# Effects of Internship on Fresh Graduates: A case study on IIT, DU students

Amit Seal Ami, Asif Imran, Alim Ul Gias and Kazi Sakib

Institute of Information Technology (IIT), University of Dhaka, Bangladesh

{amit,asif,alim,sakib@iit.du.ac.bd}

## Abstract

The aim of any curriculum is to produce industry ready students. The effectiveness of curricular activities, thus, can be measured by the performances of fresh graduates at their job sectors. To evaluate the Software Engineering (SE) syllabus, Institute of Information Technology (IIT), University of Dhaka, has taken an initiative, under the project IQAC, HEQEP, where a survey based study has been performed. The uniqueness of this SE syllabus is having a six month long internship semester inside the curriculum. Considering all the other courses and activities as traditional, the outcome of the study can fairly be considered as the effect of the Internship program. The result shows that the students having internship experiences, performed above the level of expectation from the industries.

## Core Study Team

The core team members of this study are all faculties of Institute of Information Technology (IIT), University of Dhaka (DU) and they were present during the study initiation till finalization and report generation. This team consists of four team members, who are as follows:

- Dr. Kazi Muheymin-Us-Sakib, Professor, IIT, University of Dhaka
- Alim Ul Gias, Lecturer, IIT, University of Dhaka
- Asif Imran, Lecturer, IIT, University of Dhaka
- Amit Seal Ami, Lecturer, IIT, University of Dhaka

# 1. Introduction

Bangladesh has been steadily growing its Information Technology/Information Technology Enabled Services (IT/ITES) sector steadily. There are more than 1500 registered IT/ITES providers employing more than 250,000 ICT professionals (LICT Bangladesh & BASIS Software and IT Service Catalog, 2014, Tholons Research and Estimates, 2015). Including the freelance outsourcing segment, the IT/ITES revenue generated by Bangladesh reached approximately US$600 million for the period 2013-14 and consistently growing. However, of it, only US$132.5 million came from IT/ITES export sector which is pale in comparison to RMG sector. Based on Dutch Trade Support Initiative, Bangladesh lacks visibility as an off-shoring ICT destination regardless of its forty (40) percent lesser cost. Moreover, many of the service providers are lacking in size, comprising 10-30 employees in total; making them unsuitable for handling large scale systems and projects of both domestic and foreign market in terms of development and long term services. Additionally, Bangladeshi workforce has well-developed hard skills but is believed to be less developed on soft skills (such as listening, communicating requirements) according to the interviews conducted by Nyenrode Business Universiteit (ICT Sector Study Bangladesh, 2014). As a result, even though Bangladesh has all the potential and raw elements for thriving in IT/ITES sector, it will not be able to do so unless its human resources are carefully nurtured.

One of the dominant revenue generating IT/ITES sector is customized software development. Software Engineering bridges both creativity and discipline for the creation of software. Without rigorous training from academia, one cannot engineer sustainable, maintainable, robust software; even though it is possible to develop software regardless. For both small and large scale software production, software engineering requires many roles and activities, such as requirement engineering, requirement prioritization, architecture designing, programming, quality assurance, testing and maintaining. Learning and applying all these require years of training and appropriate mindset. Considering this, many institutions and academies were setup in Bangladesh offering BSc., MSc., PGD, and short training focusing on Computer Science, Information Technology, Information and Communication Technology and Software Engineering from the 1990s. However, There is a large expectation gaps between IT graduates and Industry requirements; as found by studies conducted by BASIS, LICT and Institute of Information Technology, University of Dhaka.

Even though IT graduates are getting engineering certificates from academia, they do not get sufficient exposure or professional experience relevant to industry. Barely any graduate gain industry experience through internship or co-op works with industry before graduation. As a result, IT graduates from academia requires at least six months of training after entering industry just to get started even after average four years of academic studies. Graduates not getting assimilated in industry sometimes end up in other sectors such as administration or finance, wasting years of knowledge gathered at academia. The problem is not with the academic curriculum, however, as academic institutions focus on basic principles and engineering disciplines; but lacks more on the familiarity and knowledge of real world engineering practices and principle. The only way to mitigate this is to introduce internship and/or co-op program within the curriculum of academic institutions without substantially changing the contents of existing curriculum.

IIT, DU has addressed this issue, and integrated the six-month long internship program, where a student is sent to the industry in their 4$^{th}$ year 1$^{st}$ Semester. In that period, students stay full time in that company and work as a full time employee of that company. To ensure the integrity, students are getting paid by the companies. So that, companies will not let those resources sit idle. IIT also shares the semester credit points with the Industry, to allow the industry more control over their interns. The after effect of this process is evident with our study. We first listed the companies where our interns went. We sent survey questionnaires to the mid level employees who closely worked with our students as a peer or manager, and the HR personnel who were involved with the recruitments. We collected answers in three categories – Job specific, Communication and Interpersonal skills. We have collected about 175 survey answers and put their average results into the graph. All the results are really encouraging.

## 2. List of Employers

| List of Intern taking Companies | Website |
| --- | --- |
| AK Khan Telecom | aktelecom.net |
| Banglalink Ltd | banglalink.net |
| BDCOM Online Ltd | bdcom.com |
| Binary Quest | binaryquest.com |
| Bkash Ltd | bkash.com |
| Brac IT Systems Ltd | bracits.com |
| Brain Station 23 | brainstation-23.com |
| Data Edge Limited | data-edge.com |
| DataSoft Systems Ltd | datasoft-bd.com |
| Divine IT Ltd | divineit.net |
| Fiber@Home | fiberathome.net |
| Field Information Solutions GmBH | field.buzz |
| Four Dimension | |
| Fox AI | fox-ai.com |
| Grameen Phone Ltd | grameenphone.com |
| Graphic People Limited | graphicpeoplestudio.com |
| iBall Solutions | |
| Ibcs-primax software (bangladesh) ltd | ibcs-primax.com |
| Ice Breaker | icebreaker-bd.com |
| IDLC Finance Ltd | idlc.com |
| Jantrik Technologies Ltd | jantrik.com |
| Kaz Software | kaz.com.bd |
| Leads Corporation | leads-bd.com |
| M & H Informatics (BD) Ltd | |
| Monstar Lab Bangladesh Ltd | monstar-lab.com/Bangladesh |
| Multimedia Content and Communications Ltd | mcc.com.bd |
| Nascenia IT | nascenia.com |
| NNS Solutions | nns-solution.net |
| Orion Informatics Ltd | onirban.net |
| Panacea Systems Ltd | bdpanacea.com |
| Peoples Telecom and Information Services | ptelco.net |

| Ltd | |
|---|---|
| Pragati Software Ltd | pragatisoftware.com |
| Samsung R&D Institute Bangladesh Ltd | samsung.com |
| Secure Link Services AG | selise.ch |
| Shohoz.com | shohoz.com |
| Southtech Limited | southtechgroup.com |
| Standard Chartered Bank BD | sc.com/bd |
| Together Initiatives Ltd | i2gether.com |

## 3. Methodology of the Study

The methodology of the entire operation activities were as follows:

1. The team met by themselves and devised and documented effective strategies for data collection. One of the challenges they all focused on was to ensure maximum response rate by all groups of respondents.

2. Dates of the interview were fixed on which the teams would initiate and collect data from the respondents.

3. In this regard, the teams started to communicate and contact the respondent groups to ensure their response in the survey.

4. Popular methods of communication included emails, personal communication, social network and mobile connectivity.

5. After data entry, all data were digitized in excel which ensured that the Digital data is readily available for analysis

6. After digitization of the data, it was provided to a team of analysts who crunched the data and obtained important analytics. These analytics were then shown visually through a series of graphs, charts, etc.

7. Specific observations (quantitative) were made on the basis of the obtained data.

8. To summarize, the following strategy was used for the interview process:

| Methods | Stakeholders |
|---|---|
| Workshop | The members of the team conducted workshop where they highlighted the goal of the effort and directed formation of separate teams for data collection, data analysis and report writing. |
| Interviews | Based on the devised questionnaire, selected company personnel were approached and their responses were collected. Mostly, Human Resource department and Intern Supervisors from the companies were approached. |

| Data Entry | Specific personnel were involved to enter the data which were placed in the paper questionnaires. As a result, the data were all digitized for evaluation. |
|---|---|
| **Data Analysis** | Observations were made upon analysis of the data. Hence, the results were shown through visual mechanism. |

Table 1: Summary of the proposed methodology

Sample questionnaire is available upon request.

## 4. Following legends were used for data analysis

- 1 – Strongly Disagree/Not at all Important/Poor
- 2 – Disagree/Less Important/Fair
- 3 – Undecided/Important/Good
- 4 – Agree/Very Important/Very Good
- 5 - Strongly Agree/Extremely Important/Excellent

## 5. Employer's Feedback

All the employers were asked to mark two folds for each of the questions – one is what is the expectations from a fresh graduate and another is the performance of the IIT fresh graduates on that matter. Both are asked to mark in 5 Scale as mentioned in Section 4. The employers have emphasized that importance of the recruitment procedure is more important than academic performance of the graduates.

Figure 1 shows the academic performances of IIT fresh graduates. The first bar is showing the importance of the subject matter to recruitment and the second bar is the performance of our fresh graduates. The score differences between Academic performance to Recruitment procedure for the specific fields are given below:-

- o Job knowledge: +0.20

  Result shows that, companies want job knowledge at 3.5 level out of 5, whereas IIT students, having internship experiences, performed at 3.7 level out of 5.

- o IT knowledge: +0.018

  Result shows that, from a fresh graduate, companies want specific IT knowledge at about 3.9 level, and IIT students have a knowledge greater than 3.9.

- o Innovative knowledge: +0.20

HR departments want fresh graduates having Innovative knowledge at about 3.55 out of 5 band, and IIT fresh graduates performed at about 3.75 out of 5.

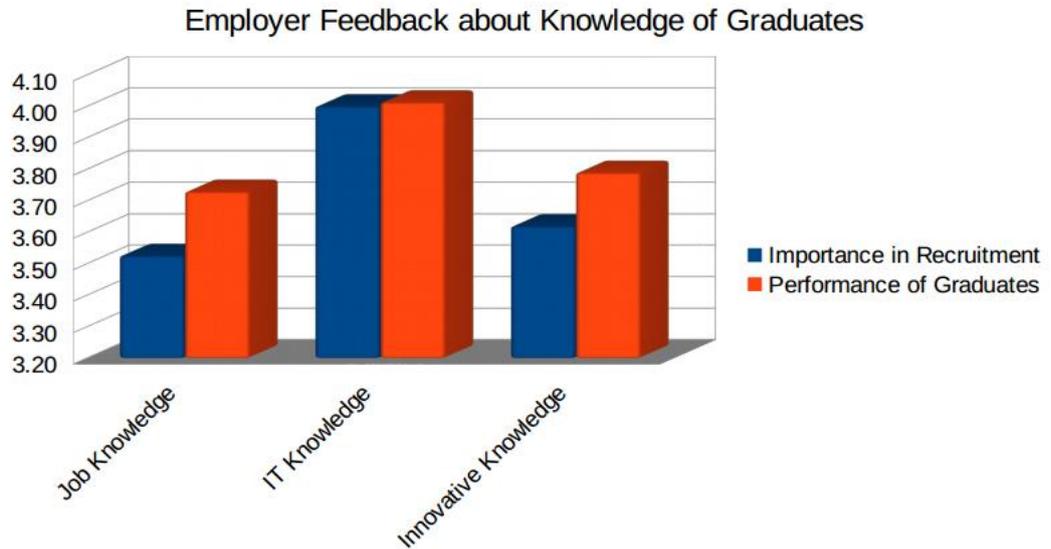

*Figure 1: Feedback from employers on the knowledge level of the graduates of IIT*

## Feedback of the Employers on Communication skills

In the employer feedback in the communication skills, the importance on the Performance of Graduates scored over the Recruitment process. The difference between the two performance attributes is shown in Figure 2 for the following criteria.

- Oral communication
- Written communication
- Presentation skills

Students are usually shy in communication, especially public university students. However, since IIT students already spent six months in the industry, their shyness drastically reduced, as shown in the figure. Mostly, they may feel homely at the job interview. That is why their oral communication is much improved, the result shows that it is at par with the industry requirements. Since students have to write reports on companies where they went as interns and also about the work they performed as interns, their technical writing ability is well above the required reference point. Students also have to present whatever they did as intern in the company premises as well as in the Institute, so their presentation skill is also commanding.

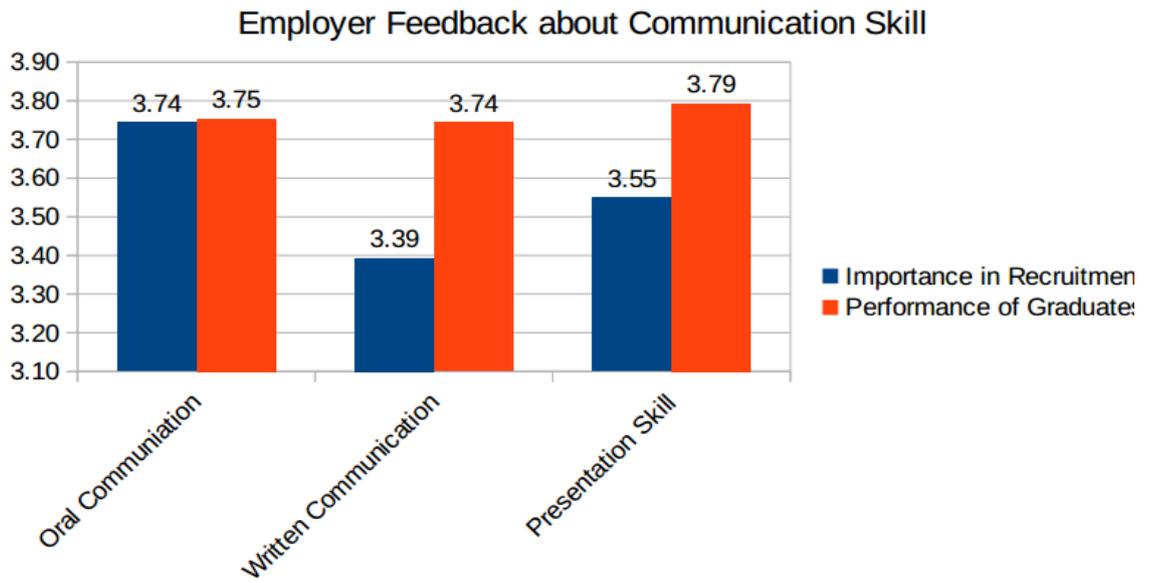

*Figure 2: Communication skills of the students of IIT*

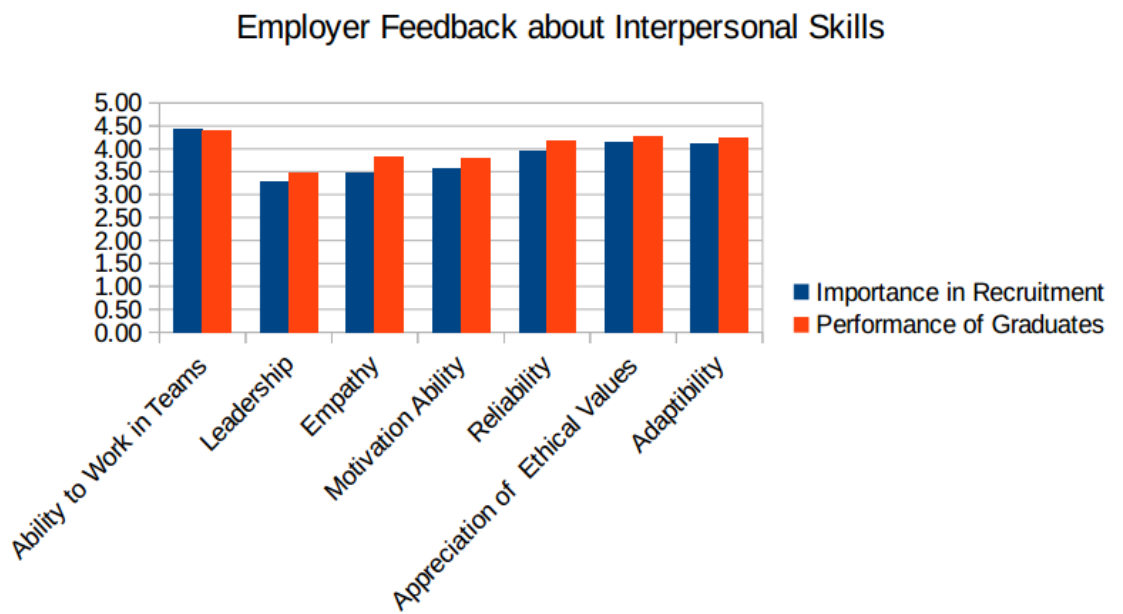

*Figure3: Work skills of the students of IIT*

## Employers Viewpoint on Interpersonal Skills

Interpersonal skills focused on leadership, empathy, motivational ability and reliability where Performance of the Graduates outscored the Recruitment quality. Figure 3 shows that in all the cases IIT graduates having intern experiences performed as par with the industry requirements.

However, Importance to work in teams attribute was decided by the Ability of Recruitment more than the Performance of the Graduates.

## Threats to Validity of the Data

- The survey participants are only based on their involvement with our students.
- Raw data is aggregated to produce the results. No data validation approach is used.
- The level required for recruitment is based on perception of the survey participants only. No benchmark or standard reference point is available for Bangladesh.
- Only companies which are previously took IIT interns are considered for the survey.

## Summary

The results are really encouraging. It is showing how important the Internship program can be for a student especially in the IT sector. However, we would like to request not to consider the numeric values as granted. Rather the perception of the professionals towards IIT fresh graduates should be highlighted.